\documentclass[smallextended]{svjour3}     % onecolumn (second format)
\usepackage{graphicx}
\advance\paperheight by-20mm
\usepackage{color}
\usepackage{amsmath}
\usepackage{amssymb}
\usepackage{savesym}     % To handle conflicts with the vec commands 
\savesymbol{vec}
\usepackage[tight,TABTOPCAP]{subfigure}   % Sub-figures
\usepackage{MnSymbol}
\usepackage{url}
\usepackage{cite}
\usepackage{hyperref}
\usepackage{marvosym}
\newcommand{\envelope}{(\raisebox{-.5pt}{\scalebox{1.45}{\Letter}}\kern-1.7pt)}
\usepackage{times}
\usepackage[LY1]{fontenc}
\papertype{generic article}

\graphicspath{{Figures/},.}

\newcommand{\ie}{{\it i.e.\ }}

\newcommand{\expec}[1]{\left<{#1}\right>}
\newcommand{\ol}[1]{\bar{#1}}

\begin{document}

% ----------------------------------------------------------------------
% TITLE and AUTHORS
% ----------------------------------------------------------------------

\title{Effects of city-size heterogeneity on epidemic spreading in a metapopulation}
%\title{Epidemic spreading in heterogeneous environments}
\subtitle{A reaction-diffusion approach}
%\titlerunning{This is the title of the article}
%\dedication{Dedicated to Professor Josef Stoer on the occasion of his 60th birthday}

% -- Authors
\author{Halvor Lund 
          \and
         Ludvig Lizana
           \and
         Ingve Simonsen
        } 
% %\authorrunning{Short form of author list} % if too long for running head

\institute{
  H. Lund \and I. Simonsen \envelope \at 
  Department of
  Physics, Norwegian University of Science and Technology (NTNU), {\\}
  NO-7491 Trondheim, Norway \\ \email{Ingve.Simonsen@ntnu.no}
  \and
  L. Lizana \envelope \at 
  Niels Bohr Institute, University of Copenhagen, Blegdamsvej 17  DK-2100 Copenhagen \O, Denmark\\ 
  Integrated Science Lab, Department of Physics, Ume{\aa} University, SE-903 34 Ume{\aa}, Sweden
  \\ \email{ludvig.lizana@physics.umu.se}
  } 

\date{Received: 1 December XXXX / 
      Accepted: 12 February XXXX / 
      Published online: XX March XXXX}

\maketitle

% ----------------------------------------------------------------------
% ABSTRACT
% ----------------------------------------------------------------------
\begin{abstract}
We review and introduce a generalized reaction-diffusion approach to epidemic spreading in a metapopulation modeled as a complex network. 
The metapopulation consists of susceptible and infected individuals that are grouped in subpopulations symbolising cities and villages that are coupled by human travel in  a  transportation network. By analytic methods and numerical simulations we calculate the fraction of infected people in the  metaopoluation in the long time limit, as well as the relevant parameters characterising the epidemic threshold that separates an epidemic from a non-epidemic phase. Within this model, we investigate the effect of a heterogeneous network topology and a heterogeneous subpopulation size distribution.  Such a system is suited for epidemic modeling where small villages and big cities exist simultaneously in the metapopulation.  We find that the heterogeneous conditions cause the epidemic threshold to be a non-trivial function of  the reaction rates (local parameters), the network's topology (global parameters) and  the cross-over population size that separates ``village dynamics'' from ``city dynamics''. 
%Finally we study in detail the effect on the epidemic threshold from the heterogeneous interactions.   
  % 
  % 
  %\subclass{65K05\and 90C35}
  % 
  % \ESM{Supplementary
  % 
  % material is available in the online version of this article
  % at \columncase{}{\\}http://dx.doi.org/10.1007/s00412-003-0244-6}
\end{abstract}

% --------------------------------------------------------------------
% MAIN TEXT
% --------------------------------------------------------------------

\section{Introduction}
Since the beginning of man, infectious diseases have played an important role in how we interact, migrate, and form societies. Diseases that are highly infectious, can soon develop into an epidemic.  A ``global epidemic'' --- often referred to as a pandemic --- is a disease that cause infection in a significant portion of a population over a large geographical area. Fortunately, most of the common infectious diseases are not lethal, but the ones that are can result in huge numbers of deaths, often significantly reducing, and even eradicating, the population in certain areas.  

% History
Historically, the world has experienced several pandemics, often with high death tolls; The plague of Justinian~(541--542AD) struck the Eastern Roman Empire and is estimated to have had a mortality of 50\% to 60\%; the Black Death of the 14th century  reduced the world population from an estimated 450 millions to between 350 and 375 millions; the tuberculosis pandemic of the 19th century killed about 25\% of the adult European population; and the Spanish flu (influenza) of 1918 is estimated to have caused the death of between 25 to 50 million people. More recently, the swine flu or the H1N1 influenza (a new variant of the Spanish flu) killed about 300,000 people within the first 12 months in 2009~\cite{dawood2012estimated}.  
In 2002, Hong Kong experienced an outbreak of the severe acute respiratory syndrome~(SARS) which,  according to the World Health Organization,  nearly became a pandemic with 8,422 cases and 916 deaths worldwide  between November 2002 and July 2003 (10.9\% fatality)~\cite{who2003}. SARS spread within weeks from Hong Kong to 37 other countries.

Needless to say, there is a huge interest in epidemic outbreaks,  spreading and containment which has attracted scientists from a number of different disciplines. 
Early pioneers in infectious disease modeling were William Hamer and Ronald Ross (Nobel laureate in medicine 1902)  who in the early twentieth century applied the law of mass action to explain epidemic behavior of common infectious fevers of childhood and malaria, respectively~\cite{ross1916application,hamer1929epidemiology}. The early work  by Kermack and MacKendrick~\cite{kermack1927contributions} on disease dissemination in a homogeneous population also deserves acknowledgment in this context. Since then numerous works have appeared of  increasing  complexity and detail.
One set of models concern spatial dynamics   dealing with epidemic spreading over geographic areas~\cite{mollison1977spatial}. Examples range from forest-fire models~\cite{rhodes1997critical} to spatial immunization patterns where multiple diseases compete for their hosts~\cite{sneppen2010minimal},  and explicit simulation of 85 million people including households, schools and workplaces in Thailand~\cite{ferguson2005strategies},  a region of high interest due to the  avian H5N1 epidemic.  

Another class  is  metapopulation models. Such models are based on the observation that large human populations are typically divided into a network of smaller subpopulations such as cities or villages.  Crudely speaking, a metapopulation consists of a group of connected subpopulations. Between the subpopulations there is an ongoing exchange of people, some of which are carriers of a disease and may infect previously unaffected subpopulations~\cite{colizza2007reaction,vespignani2011modelling,colizza2007invasion,grenfell2001travelling,earn2000simple}.  The dynamics in each ``node''  is often assumed to be governed by mass-action kinetics, or sometimes stochastic (relevant for very low population densities). Obviously, the way in which the subpopulations are interconnected has a significant effect on how the disease spreads through the metapopulation. Particular emphasis has therefore, on good grounds,  been paid to the impact of airline travel  using explicitly, or features of, the world's airline transportation network~\cite{brockmann2009human,hufnagel2004forecast,colizza2006role,colizza2007modeling,ferguson2006strategies}. Recent efforts also try to improve the description of human mobility patterns such as recalling individuals' geographic origins~\cite{brockmann2006scaling,balcan2011phase}.

A third model class, albeit related to the former, is network models that account for how humans actually meet~\cite{newman2002spread,pastor2001epidemic,kuperman2001small}. In metapopulation models it is often assumed that the population within a single subpopulation is well mixed, which means that every individual is equally likely to spread the disease to any other member of the population. This is of course not generally true. For example, in large cities any individual only meets a very small fraction of the total population. There is, however, frequent encounters between those belonging to the same social network. 

There are also interesting works focusing on the situation where humans are not the main carriers of the disease. Examples include Bubonic plague carried by rats~\cite{keeling2000metapopulation} and  water-borne cholera epidemic in a South African river network~\cite{pinto2012locating}.

In this paper, we use a metapopulation description to model disease spreading in a network of cites and villages (nodes) of varying population sizes which are interconnected via a transportation network. The infection model we assume is the well-established susceptible-infected-susceptible model~(Sec.~\ref{sec:SIS}) where individuals can be in either a susceptible or an infected state and going from one state to the other is  controlled by rates. For an isolated city or village, the fraction of infected individuals in the population in the final (stationary) state is fully determined by these rates. However,  when the cities and villages in addition are exchanging people with each other, for instance on a transport network, the situation may be much more involved. Under such conditions, the final state is not only determined by the rates, but  also  depends on  structural features of the network~\cite{colizza2007reaction,vespignani2011modelling}.

The final state of the metapopulation can be either epidemic or non-epidemic, meaning that a fraction of the population is infected (or not). The separation between  these two phases (in parameter space) is known as the epidemic threshold. In this study we go beyond previous works and address the importance of city-size heterogeneity on the epidemic threshold. The size of the city or village is an important aspect because it sets the limit on how large fraction of the population a single individual actually meet within a given time period. The well-mixed approximation works poorly for a large city like New York  whereas for a small countryside village it could be adequate. We therefore let the infection dynamics at each node change depending on how many individuals are present at any given time. In previous models, the network's topology only influenced how people, and thus the disease,  spread between the nodes. Here we in addition let the dynamics adapt to the local instantaneous population density, which  means that the network structure influence the disease dynamics in two ways. First, how individuals move, and second, how they interact on the nodes. The latter part is in practise achieved by introducing a population dependent infection rate that interpolates between the two limiting cases of ``the big city'' for which the infection rate is dropping with increasing population, and ``the small village'' where the infection rate remains independent of population size. Within this setting, we derive new analytic expressions for the epidemic threshold of a general network which interpolates between previously known cases.

\medskip 
This paper is organized as follows; Section~\ref{sec:SIS} is devoted to detailing the infection model that will be assumed. Here also the behavior of this model in homogeneous space is discussed. We then present how we model the transportation of individuals between the cities and villages~(Sec.~\ref{Sec:Diffusion}) before we in Sec.~\ref{Sec:Reaction-Diffusion} present a unified reaction-diffusion model that combines the two phenomena from the previous two sections. 
Section~\ref{Sec:Reaction-Diffusion} also presents analytic results for the epidemic threshold and compare them to simulations.  
The conclusions from  this study are presented in Sec.~\ref{Sec:Conclusions}.

\section{The susceptible-infected-susceptible epidemic model}
\label{sec:SIS}

The model that we will focus on in this work is the classic susceptible-infected-susceptible~(SIS) epidemic model~\cite{Book:Keeling-2008}. This model  is chosen since  it is relevant for an important group of infectious diseases. Moreover, it has been studied extensively in the literature, and it can often be given an analytic treatment. 

In the standard SIS model there exists two types of individuals --- the individuals that are susceptible to an infection~(labeled S below), and those individuals that already are infected and can infect others (labeled I).  If an infected individual meets an uninfected individual, he will transmit the disease at rate $\beta$. On the other hand, an infected individual can recover from the illness at rate $\mu$ and thereby again becoming susceptible to the disease.  Formally, the reactions of the SIS model can be written in the form
\begin{subequations}
  \label{eq:SIS-reactions}
\begin{align}
  S+I & \overset{\beta}{\longrightarrow} 2I, \\
  I   & \overset{\mu}{\longrightarrow} S,
%  \susceptible + \infected  & \overset{\beta}{\longrightarrow}  2 \infected, \\
%                 \infected  & \overset{\mu}{\longrightarrow}  \susceptible,
\end{align}
\end{subequations}
where we explicitly have indicated the \emph{infection} (reaction) rate $\beta\geq 0$, and the \emph{recovery} (healing) rate $\mu\geq 0$.
The SIS model is the prototype for a disease spreading model where individuals that have recovered from a disease will automatically be susceptible to re-infection. On the other hand, if individuals that recover from the disease become immune to it, the classic  (prototype) epidemic model describing this situation is the related susceptible-infected-removed~(SIR) model~\cite{Book:Keeling-2008}. This latter model, however, will not be discussed further in this work.

The first property to note for the dynamics of the SIS model~\eqref{eq:SIS-reactions}  is that it conserves the total number of individuals, or in the language of physics, the total number of ``particles''. This number we denote by
\begin{align}
  \label{eq:total_no_particles}
  N=N_S(t)+N_I(t),
\end{align}
where $N_p(t)$ is the total number of individuals of type $p=I,S$ at time $t$. Note that in writing Eq.~\eqref{eq:total_no_particles}, we have explicitly indicated a time-dependence for $N_p(t)$ since the number of individuals of type $p$ may fluctuate in time even if the total number of individuals $N$ in the system remains constant at all times.  The second feature of the SIS dynamics is the active role played by the infected individuals (the ``$I$ particles''); if they have disappeared completely at time $t_0$, {\it i.e.} $N_I(t_0)\equiv 0$, there is no way in the model to reintroduce the infection and, therefore, the population remains uninfected for all later times $t\geq t_0$.

\subsection{SIS dynamics in homogeneous space}
\label{Sec:SIS-dyn}
In order to understand the dynamics of the SIS model, we will start by studying it in homogeneous (continuous) space under the assumptions that susceptible and infected individuals are well mixed. Based on reaction~\eqref{eq:SIS-reactions} we consider the model (in discrete time)
\begin{subequations}
  \label{eq:reactions_eq}
\begin{align}
  \label{eq:reactions_eq_A}
   N_{S}(t+1) &=   N_{S}(t) 
                + \mu N_{I}(t) 
                - \beta \frac{ N_{S}(t)N_{I}(t) }{N_\times + N }, \\
    \label{eq:reactions_eq_B}
    N_{I}(t+1) &=   N_{I}(t) 
                - \mu N_{I}(t) 
                + \beta \frac{ N_{S}(t)N_{I}(t) }{N_\times + N },
\end{align}
\end{subequations}
where the second and third term on the right-hand-side of these equations are associated with recovery and infection, respectively. 
For later convenience, we have in writing Eq.~\eqref{eq:reactions_eq} introduced a characteristic population size, $N_\times$, so that that the effective infection rate is
\begin{align}
  \label{eq:effective_reaction_rate}
  \beta' = \frac{\beta}{N_\times+N} 
         =  
         \begin{cases}
           \frac{\beta}{N_\times}, & N \ll N_\times; \qquad \mbox{"the small village"}\\
           \frac{\beta}{N},       & N \gg N_\times;  \qquad \mbox{"the big city"}
         \end{cases}.
\end{align}
From Eq.~\eqref{eq:effective_reaction_rate} it follows that in the limit $N \ll N_\times$, the infection rate $\beta'$ is \emph{independent} of the total population size $N$. This situation is typical for a smaller community where one, within a rather short period of time, may be ``exposed'' to almost the entire  population of the community. For instance, this may happen during a funeral, a music gathering or a Christmas party. For this reason, we will below refer to this type of interaction being of  ``the small village'' type. In big cities, on the other hand,  the probability per unit time to meet, and therefore potentially infect, another  individual is reduced as the population of the city increases. In this situation we have chosen to model the effective infection rate with Eq.~\eqref{eq:effective_reaction_rate} so that $\beta'\sim N^{-1}$ in the limit where $N\gg N_\times$.

\subsection{Stationary state}
Under stationary conditions $N_I(t)$ and $N_S(t)$ are independent of time $t$, which we denote by $\bar{N}_I$ and  $\bar{N}_S$, respectively. From Eqs.~\eqref{eq:reactions_eq} it follows that in a stationary state, the following relation must be satisfied
\begin{align}
  \label{eq:condition}
  \mu \bar{N}_I =  \beta \frac{\bar{N}_S \bar{N}_I }{ N_\times + N},     %
  %\beta' \bar{N}_S \bar{N}_I,
  %\frac{\bar{N}_S}{N} = \frac{\mu}{N \beta'} = \frac{\mu}{\beta} \left( 1+\frac{N_\times}{N} \right). 
\end{align}
with the additional constraint (see Eq.~\eqref{eq:total_no_particles}) that $\bar{N}_S + \bar{N}_I=N$. Equation~\eqref{eq:condition} is trivially fulfilled for $\bar{N}_I=0$ (and $\bar{N}_S=N$), meaning that the disease has died out and will never reappear. However, Eq.~\eqref{eq:condition} is also satisfied if $\bar{N}_S=\mu/\beta'$  with the requirement that $\bar{N}_I=N -\bar{N}_S\geq 0$.
The (non-trivial) condition reads %for $\mu$ and $\beta$ is
\begin{align}
  \label{eq:condition-2}
  \bar{N}_I
       %=  N-\bar{N}_S
       =  N \left[ 1- \frac{\mu}{\beta} \left( 1+ \frac{N_\times}{N} \right) \right]
      > 0,
\end{align}
which  holds if
\begin{align}
  \label{eq:epidemic_treshold}
  \frac{\beta}{\mu} >  1+ \frac{N_\times}{N}.
\end{align}
This means that an ``epidemic phase'' exists ($\bar{N}_I>0$) in the long time limit when individuals are infected faster than they recover .  In all other cases, the only acceptable solution to Eq.~\eqref{eq:condition} is the  non-epidemic solution  $\bar{N}_I\equiv 0$ and $\bar{N}_S\equiv N$. Since the boarder between these phases occurs when  $\beta/\mu =  1+ N_\times/N$, this point in parameter space is referred to as the {\em epidemic threshold}~\cite{Book:Keeling-2008}.   

%Then, by solving the rate equations for the number of $A$ and $B$ individuals, it is readily shown that an epidemic threshold exists at $\beta/\mu=1$; when $\beta/\mu<0$ individuals will recover quicker than they are infected with the result that the infection that initially existed in the population will die out in the long time limit, and for  $\beta/\mu<0$ an epidemic phase does exist for which a sustainable faction of the population remains infected~\cite{Book:Keeling-2008}. T

%\section{Complex network of population centers}

%We are all very aware of the fact that the assumption of a homogeneous and well mixed population make in the previous section is incorrect. Instead, throughout history we have formed local communities and today they range in size from small villages to mega-cities. 

\section{Migration within a metapopulation}
\label{Sec:Diffusion}
\begin{figure}[tbp] %[ht]=here [t]=top [b]=bottom [p]=separat side
\centering
\includegraphics[width=0.98\textwidth]{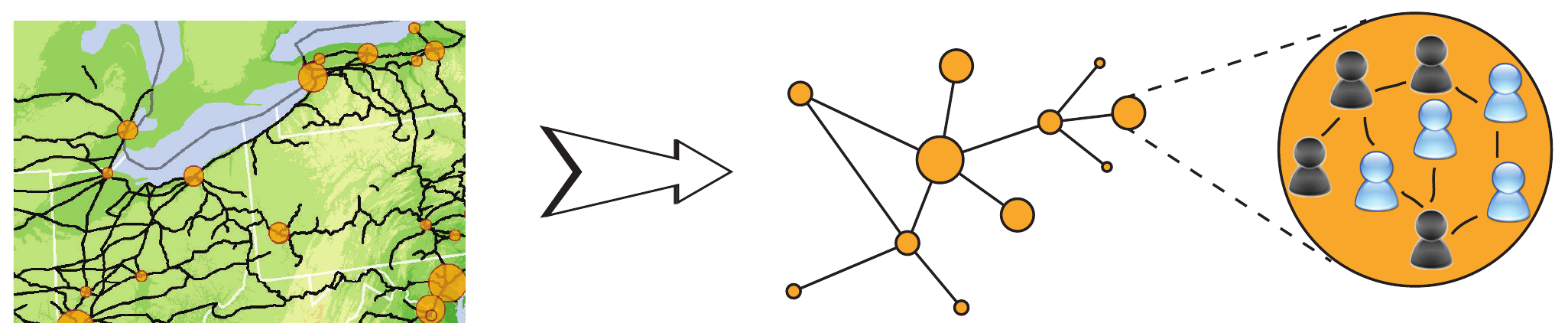}
\caption[]{Schematics of the metapopulation model used in this paper. A real network of subpopulations is mapped onto a complex network. Individuals are distributed over the nodes of the network where the susceptible (light blue men) and infected (dark) individuals interact on the nodes according to the SIS dynamics (Eq. ~\eqref{eq:reactions_eq}), and travel between the subpubulations (nodes) by diffusion.}
\label{fig:Schematics}
\end{figure}

In the previous section, we considered the dynamics of the SIS model under the assumption of an isolated, homogeneous  well mixed population. It was shown that an epidemic phase may exist. However, humans are not uniformly distributed over the surface of the earth. Instead, throughout history, we have had the tendency to form local communities (``subpopulations''), {\it i.e.} areas where the population density is higher than in its surroundings.  Today the population density of the world is rather heterogeneous, and the communities range in size from small villages, consisting of a few tens of people, to mega-cities or urban areas where many millions of people are living within a rather small area. A group of such spatially separated populations that are coupled to each other in one way or the other is called a \emph{metapopulation} (Fig.~\ref{fig:Schematics}). Within one  local community of a metapopulation  one may argue that the SIS interactions~\eqref{eq:SIS-reactions} may take place essentially within homogeneous space,  so that the epidemic threshold for that community is given by Eq.~\eqref{eq:epidemic_treshold}. However, for this equation to be correct one is required to neglect the exchange of people between the various communities of the metapopulation. Today we travel more than ever before; we visit neighboring villages and cities, we travel to further away regional centers, and do international travels to far away destinations for business or pleasure. Hence, the effect of the ever increasing human travel patterns~\cite{brockmann2009human,hufnagel2004forecast,colizza2006role} is that the  population centers are coupled, something that in some cases (as we will see below) influences significantly the  epidemic threshold.

% --- Figure (start) -----------------------------------------------------

% --- Figure (end) -----------------------------------------------------

% --- single world society
In our model, the nodes of the network represent  population centers and the links connecting them correspond to the human traveling routes between such centers~(Fig.~\ref{fig:Schematics}). An integrated world will be assumed; by this we mean that any node of the network can be connected to any other node by following the links of the network. Technically this is to say  that the network is fully connected and therefore consists of one giant component~\cite{Newman}. We let the network consist of $V$ nodes (or vertices), and let $N_{p}(i,t)$ denote the number of individuals of type $p=I,S$ at population center (or node) $i=1,2,\ldots,V$ at time $t$. Thus, the population of node  $i$ at time $t$ becomes $N(i,t)=N_S(i,t)+N_I(i,t)$, and by summing over $i$, the total $p$-population becomes $N_p(t)=\sum_iN_p(i,t)$. Therefore, the total population of the whole network, independent of the type of individual, can be written as  the time-independent constant $N=N_S(t)+N_I(t)$. Furthermore, the topology of the complex network is described by the so-called adjacency matrix ${\mathbf W}$~\cite{Newman,SantoCommunities}. The components of this matrix, $W_{ij}\geq 0$, describe the weight or capacity of the traveling route from population center $j$ towards population center $i$. To keep matters simple, we will assume that the links are symmetric: the capacity of the route $j\rightarrow i$ is equal to that of $i \rightarrow j$ (\ie $W_{ij}=W_{ji}$). Moreover, $W_{ij}=0$ means that no direct migration (or travel) is possible from  $j$ towards $i$.  

To obtain and use a detailed and accurate description of the traveling patterns of man is a complicated and challenging matter~\cite{brockmann2009human}, and it is outside the scope of this work. Here we will take a simplistic view and use a stochastic migration model where no assumptions are made on how people migrate. Individuals ``leaving'' population center $i$ will choose a route from center $i$ to $j$ with a probability that is proportional to the capacity $W_{ji}$. In the next time step, the same procedure is repeated again. This simple migration model is essentially what in physics is called a random walk or discrete diffusion process on the complex network~\cite{Book:Redner-2001,SimonsenInternet,SimonsenDiff,SimonsenDiff1,SimonsenCascading}.

If the susceptible and infected individuals travel according to two independent diffusive processes, it follows from the conservation of ``particles''~\cite{SimonsenInternet,SimonsenDiff,SimonsenDiff1} that the pure migration processes ($\mu=\beta=0$) satisfy
\begin{subequations}
  \label{eq:diffusion}
\begin{align}
  \label{eq:diffusion-A}
  N_p(i,t+1) = N_p(i,t) + D_p \sum_{j=1}^V T_{ij} N_p(j,t) - D_p N_p(i,t).
\end{align}
Here $D_p$ denotes the ``diffusion constant'' for $p$-type individuals,  with $0\leq D_p \leq 1$, or more precisely, the traveling probability. 
Moreover, in writing Eq.~\eqref{eq:diffusion-A} we have defined a transfer matrix, ${\mathbf T}$ with elements 
\begin{align}
  T_{ij} = \frac{W_{ij}}{\omega_j},     %{\sum_iW_{ij}},   %{\omega_j},
\end{align}
where the total outgoing capacity from node $j$ is denoted
\begin{align}
  \omega_j = \sum_{i=1}^V W_{ij}.
\end{align}
\end{subequations}

Under the assumption $\mu=\beta=0$, the number of $S$- and $I$-individuals are separately conserved at any time, and the population sizes in the stationary states of the diffusive migration process are given by
\begin{align}
  \label{eq:Stationalry_state_diffusion}
  \bar{N}_p(i) \propto \omega_i, \qquad p=S,I.
\end{align}
This result is readily obtained  by substituting the above expression into Eq.~\eqref{eq:diffusion} and using that the adjacency matrix is symmetric. The implication of  equation~\eqref{eq:Stationalry_state_diffusion} is that the size of city/village $i$ in the stationary state is proportional to the total migration capacity $\omega_i$ of that population center, and it is independent of what the diffusive constant might be. As a result of Eq.~\eqref{eq:Stationalry_state_diffusion}, a separate subpopulation size distribution of villages and cities will not be put explicitly into the model. Instead such information enters implicitly  {\it via} the distribution of weights $\omega_i$, at least it does so for the pure diffusion process. Equation~\eqref{eq:Stationalry_state_diffusion} appeals to intuition because larger cities will have transportation systems of larger capacity than smaller cities or villages, in order to serve the larger population. In passing we note that diffusion on complex networks, as described by Eq.~\eqref{eq:diffusion}, has previously been used to detect network clusters (communities)~\cite{SimonsenInternet,SimonsenDiff,SimonsenDiff1,SantoCommunities}, and for the cascading failures of complex networks due to overloading~\cite{SimonsenCascading}, to mention a few examples.

\section{A reaction-diffusion approach to epidemic spreading in a metapopulation}
\label{Sec:Reaction-Diffusion}

We will now combine the two phenomena from the previous two sections;  infection (reactions) and migration (diffusion). To facilitate future discussion,  we introduce the average nodal population density of the network, $\rho$, defined as the average number of individuals per node (or vertex)
%\begin{subequations}
\begin{align}
  \rho = \frac{N}{V}, %= \rho_S(t)+\rho_I(t), 
\end{align}
where it is recalled that $V$ is the number of nodes in the network. In terms of population density, Eq.~\eqref{eq:total_no_particles} becomes\begin{align}
  \rho =\rho_S(t)+\rho_I(t), \qquad  \rho_p(t) = \frac{N_p(t)}{V}, \quad p=S,I.
  \label{eq:rho}
\end{align}
%\end{subequations}
The infection and migration processes of individuals in a metapopulation can now be simulated, given some initial values for $N_S(i,0)$ and $N_I(i,0)$, a realization of the network, and values for the model parameters. The  equations on which such simulations are based are obtained by combining Eqs.~\eqref{eq:reactions_eq} and \eqref{eq:diffusion} so that the relevant set of equations is of  reaction-diffusion type. Motivated by the key question in epidemiology studies, we focus on the long time behavior of this reaction-diffusion system. Therefore, our primary interest will be the average nodal density of infected individuals in the stationary state of the system, $\bar{\rho}_I$. The corresponding density of susceptibles is readily obtained as $\bar{\rho}_S=\rho-\bar{\rho}_I$  and will thus not be explicitly given.  We recall from the preceding discussion that a non-trivial result $\bar{\rho}_I>0$ requires that initially $N_I(i,0)>0$ for at least one node in the network. Therefore, it will from now onward be assumed that $N_I(i,0)>0$  for some $i$'s. Moreover, if $D_I$ is zero, there is no way that an initial infection can be transmitted in our model to other parts of the metapopulation. Since such a ``local infection''  is not a very interesting and/or realistic situation, we will require that $D_I>0$.

\subsection{Numerical results} \label{sec:numerical_results}
\begin{figure}[tbp] %[ht]=here [t]=top [b]=bottom [p]=separat side
\centering
\subfigure[Scale-free network, $\gamma = 3.0$]{
\includegraphics[width=0.48\textwidth]{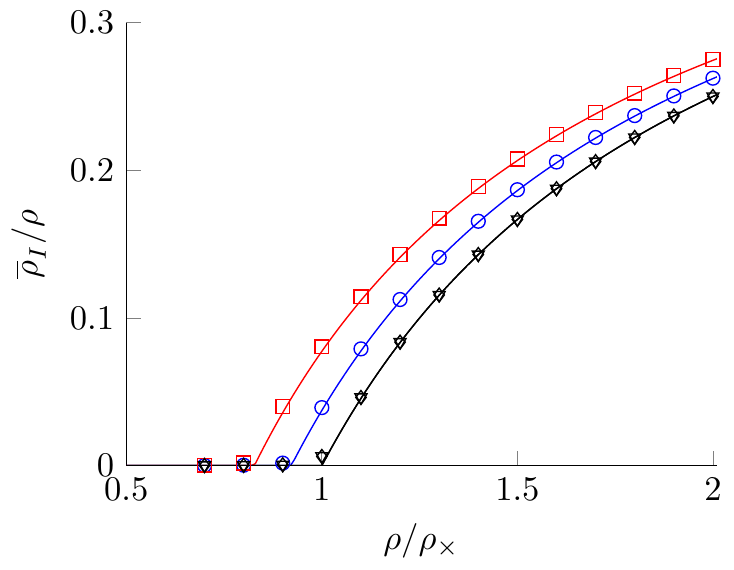}  % Putter inn eps-fila
\label{fig:statTMG3}
}
%\qquad
\subfigure[Scale-free network, $\gamma = 2.5$]{
\includegraphics[width=0.48\textwidth]{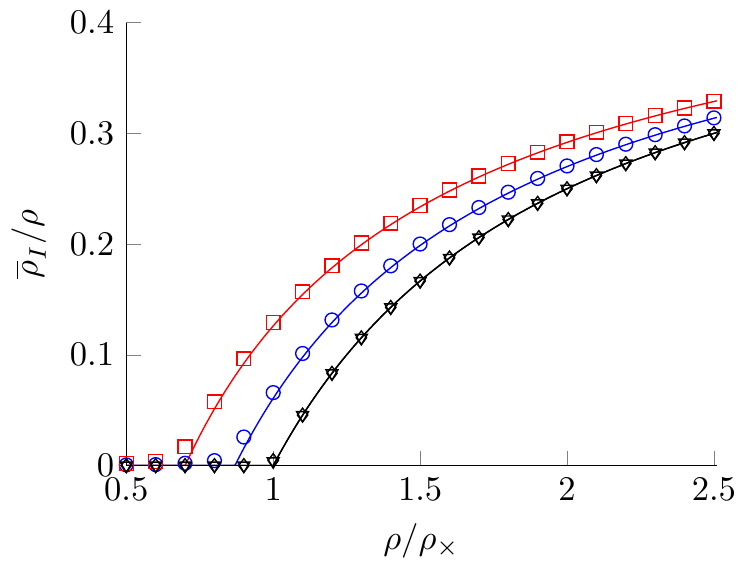}  % Putter inn eps-fila
\label{fig:typeM3}
}
\caption[]{The nodal density of infected individuals in the stationary state ($\bar{\rho}_I$) obtained by simulations (open symbols) or a mean-field approximation (solid lines). The underlying networks were assumed to have a scale-free degree distribution: $p(k)\sim k^{-\gamma}$ with $\gamma=3.0$ and $\gamma=2.5$. The parameters used were $D_I = 1$ and $\beta/\mu = 2$, and the various opens symbols labeled in terms of ($V$, $D_S$) are: Circles $\medcircle$ ($10^4$, 1), squares $\medsquare$ ($10^2$, 1), diamonds $\meddiamond$ ($10^4$, 0) and triangles $\medtriangledown$ ($10^2$, 0). We point out that the curves corresponding to the diamonds and triangles (black) fall onto the same line since $D_S=0$. The simulations were based on a (sequential) combination of  Eqs.~\eqref{eq:reactions_eq} and \eqref{eq:diffusion}  and the analytic results were obtained from Eqs.~\eqref{eq:CriticalPoint_DA_0} and \eqref{eq:CriticalPoint_DA_1}. }
%
%\caption[]{Simulation results for type M reactions, with $D_I = 1$ and $\beta/\mu = 2$. Solid lines represents the analytic result (Eqs.~\eqref{eq:CriticalPoint_DA_0} and \eqref{eq:CriticalPoint_DA_1}) while the symbols are the simulation results. Each symbol with properties ($V$, $D_S$): Circles $\medcircle$ ($10^4$, 1), squares $\medsquare$ ($10^2$, 1), diamonds $\meddiamond$ ($10^4$, 0) and triangles $\medtriangledown$ ($10^2$, 0).}
%
\label{fig:typeM}
\end{figure}

Figure~\ref{fig:typeM} shows simulation results (open symbols) for the stationary state of the (scaled) nodal density of infected individuals, $\bar{\rho}_I/\rho$,  {\it vs.} the average nodal population density $\rho/\rho_\times$ (with $\rho_\times=N_\times/V$) assuming that $\beta/\mu=2$. 
The underlying  network connecting the subpopulations was assumed to be non-weighted  and of  scale-free type, $p(k)\sim k^{-\gamma}$, characterized by the exponents  $\gamma=3.0$~(Fig.~\ref{fig:statTMG3}) and $\gamma=2.5$~(Fig.~\ref{fig:typeM3}). Non-weighted simply means that the population density at given node is proportional its number of incoming links (see Eq.~\eqref{eq:Stationalry_state_diffusion}). The simulation results shown in Fig.~\ref{fig:typeM} (as solid lines) were obtained under the assumption of treating $\rho_p(i,t)$ as a continuous variable to save computation time, and the networks were generated by the Molloy-Reed algorithm~\cite{Molloy}. A stochastic simulation approach similar to  that being used in Ref.~\cite{colizza2007reaction} was also implemented. It was found to give equivalent results except for a higher expenditure of computer time. 
% --- Figure (start) -----------------------------------------------------

% --- Figure (end) -----------------------------------------------------

It is clearly seen from Fig.~\ref{fig:typeM} that for sufficiently low nodal population density, $\rho$, the initial infection will eventually die out and leave the whole population uninfected. Under such conditions, there is no potential risk of the initial infection developing into a large scale epidemic (pandemic). The same simulation also shows that there exists a critical population density, $\rho=\rho_c$, above which an average nodal density will result in a non-zero fraction of the population being infected even in the long time limit. This means that for $\rho>\rho_c$, we are in an epidemic-phase defined by $\bar{\rho}_I>0$. Such a behavior is characteristic for what in physics is known as a {\em phase-transition}~\cite{stanley1987introduction}, for instance, observed when water goes from a solid to a liquid phase with increasing temperature.  

In passing, it should be mentioned that in the simulations performed to produce Fig.~\ref{fig:typeM}, it was assumed that reaction and diffusion were sequential processes,  not simultaneous. This is the same assumption made in the original work~\cite{colizza2007reaction,vespignani2011modelling}. However, in a subsequent study~\cite{Saldana} it was pointed out that the continuous-time limit of the discrete process used in \cite{colizza2007reaction,vespignani2011modelling} (and in this work) is not well defined if the reaction-diffusion process were sequential, and the author proposed a formulation for which a continuous-time limit could be obtained. However, doing so renders the formulation more complex with rather similar conclusions. So, we have for the sake of clarity  assumed a sequential  two-step reaction-diffusion process even if we are well aware of its shortcoming when taking the continuous limit.

%\subsection{The Critical point for the epidemic phase}
\subsection{Analytical treatment of the model}

The critical average nodal density $\rho=\rho_c$ is of great importance since it separates the non-epidemic from the epidemic phase, or in other words, defines the model's epidemic threshold. The purpose of this section is to obtain an analytic expression for this critical population density in terms of the model parameters. To this end, we again follow~\cite{colizza2007reaction,vespignani2011modelling} and impose a mean-field approximation where all nodes of the same degree $k$ are considered equivalent. We introduce the nodal population density
\begin{align}
  \rho_{p,k}(t) &= \frac{N_{p,k}(t)}{V_k},
\end{align}
where $N_{p,k}(t)$ denotes the total number of $p=S,I$-type individuals at nodes of degree $k$ in the network, and $V_k$ represents the number of nodes of degree $k$. Just like  in the simulations we assume a two-step reaction-diffusion process with the reaction step being the first of the two. Since the reaction step is local to a node, the corresponding equations for the nodal densities $\rho_{p,k}(t)$ are readily obtained from Eq.~\eqref{eq:reactions_eq} after a division by $V_k$. The diffusion step, however, needs further discussion because  the formalism used in Sec.~\ref{Sec:Diffusion}, involving  transfer matrices, is no longer optimal.  For the migration of individuals of type $p$  into and away from a node of degree $k$, we obtain the following nodal density
\begin{align}
  \rho_{p,k}(t+1) = \rho_{p,k}(t) - D_p \rho_{p,k}(t) + k \sum_{k'} p(k'|k)\frac{D_p}{k'} \rho_{p,k'}(t),
  % \frac{D_p k}{\left<k\right>} \sum_{k'} p(k') \rho_{p,k'}(t) 
  \label{eq:diffusion2}
\end{align}
that follows from the conservation of nodal density of individuals of type $p=S$ or $I$. The physical interpretation of Eq.~\eqref{eq:diffusion2} is as follows: the first term on the right-hand-side is the density that was already present at nodes of degree $k$ at time $t$; the second term is associated with the diffusion of a fraction $D_p$ of this density {\em away} from nodes of degree $k$; and the last term represents the diffusion \emph{into} nodes of degree $k$ from neighboring nodes of degrees $k'$. In writing  Eq.~\eqref{eq:diffusion2}, we have introduced the conditional probability density, $p(k'|k)$, for a node of degree $k$ being connected to a node of degree $k'$. If the network is assumed to be uncorrelated, this probability is known to be~\cite{NewmanStrogatz} % $p(k'|k) = p(k') k' / \expec{k'}$
\begin{align}
  p(k'|k) = p(k')  \frac{k'}{\left<k\right>}, 
  \label{eq:ConditionalProb}
\end{align}
where $p(k)$ is the degree distribution and $\left<k\right> = \sum_k k\, p(k)$ denotes the average degree. By substituting Eq.~\eqref{eq:ConditionalProb} into Eq.~\eqref{eq:diffusion2} and performing the sum over $k'$ in the resulting equation, one is led to  
\begin{subequations}
\begin{align}
  \rho_{p,k}(t+1) = \rho_{p,k}(t) - D_p \rho_{p,k}(t) + \frac{D_p k}{\left<k\right>} \rho_p(t),
  % \frac{D_p k}{\left<k\right>} \sum_{k'} p(k') \rho_{p,k'}(t) 
  \label{eq:diffusion3}
\end{align}
with
\begin{align}
  \rho_p(t) = \sum_{k'} p(k') \rho_{p,k'}(t).
\end{align}
\end{subequations}
%
%To obtain the final set of coupled equations for $\rho_{S,k}(t)$  and $\rho_{I,k}(t)$ for the (sequential) reaction-diffusion process, we proceed by replacing  the densities $p_{p,k}(t)$ appearing on the right-hand-side of Eq.~\eqref{eq:diffusion3}, by the right-hand-side of Eq.~\eqref{eq:reactions_eq} after dividing it by $V_k$ and evaluating it at time $t-1$. The result is
% 
To obtain the final set of coupled equations for $\rho_{S,k}(t)$  and $\rho_{I,k}(t)$  we proceed by first replacing $t$ by $t+1$ in Eq.~\eqref{eq:diffusion3} and then replacing  the densities $\rho_{p,k}(t+1)$ appearing on the right-hand side of the resulting equation by the right-hand side of Eq.~\eqref{eq:reactions_eq} after dividing it by $V_k$. The result is 
\begin{subequations}
\label{eq:reacdiff2}
\begin{align}
	\partial_t \rho_{S,k}(t) 
            =& -\rho_{S,k}(t) + (1-D_S)\left[\rho_{S,k}(t) + \mu \rho_{I,k}(t) - \beta \Gamma_k(t)\right] \notag \\
	     & + \frac{D_S k}{\left<k\right>} \left[\rho_S(t) + \mu \rho_I(t) - \beta \Gamma(t)\right] \; , \label{eq:reacdiffA2} \\
	\partial_t \rho_{I,k}(t) 
            =& -\rho_{I,k}(t) + (1-D_I)\left[(1-\mu)\rho_{I,k}(t) + \beta \Gamma_k(t)\right] \notag \\
	     & + \frac{D_I k}{\left<k\right>} \left[(1-\mu)\rho_I(t) + \beta \Gamma(t) \right] \; , \label{eq:reacdiffB2}
\end{align}
where the  reaction kernel 
\begin{align}
  \label{eq:Gamma_k}
  \Gamma_k(t) =& \frac{ \rho_{S,k}(t) \rho_{I,k}(t) }{ \rho_\times + \rho_k }, 
 \end{align}
and its average 
\begin{align}
  \label{eq:Gamma_k_average}
  \Gamma(t)   =& \sum_k p(k)\Gamma_k(t), 
\end{align}
\end{subequations}
were  introduced. In writing Eq.~\eqref{eq:reacdiff2} we have introduced the shorthand notation $\partial_t \rho_{S,k} = \rho_{S,k}(t+2) - \rho_{S,k}(t)$ (and similarly for $\rho_{I,k}$)%
\footnote{The form of these equations reflects the assumption of a sequential process where a reaction step is taking place at every node before a diffusion step. They are similar to those presented in the supplementary information accompanying\cite{colizza2007reaction,vespignani2011modelling} with the exception that here a more general reaction kernel has been used.}.

In order to obtain the stationary state of Eqs.~\eqref{eq:reacdiffA2} and \eqref{eq:reacdiffB2}, we require that $\rho_{p,k}(t) \equiv \bar{\rho}_{p,k}$ (with $p=S,I$) is independent of time so that
\begin{subequations}
\begin{align}
	\bar{\rho}_{S,k} &= (1-D_S)\left[\bar{\rho}_{S,k} + \mu \bar{\rho}_{I,k} - \beta \bar{\Gamma}_k\right] + \frac{D_S k}{\left<k\right>} \left[\bar{\rho}_S + \mu \bar{\rho}_I - \beta \bar{\Gamma}\right] \; , \label{eq:statA} \\
	\bar{\rho}_{I,k} &= (1-D_I)\left[(1-\mu)\bar{\rho}_{I,k} + \beta \bar{\Gamma}_k\right] + \frac{D_I k}{\expec{k}} \left[(1-\mu)\bar{\rho}_I + \beta \bar{\Gamma} \right],\label{eq:statB} 
\end{align}
\label{eq:stat}
\end{subequations}
where $\bar{\Gamma}_k$ is given by Eq.~\eqref{eq:Gamma_k}, but with $\rho_{p,k}(t)$ replaced by $\ol{\rho}_{p,k}$ and $\bar{\Gamma}$ is defined as an average over $\bar{\Gamma}_k$ similar to Eq.~\eqref{eq:Gamma_k_average}.
Following the strategy from Ref.~\cite{colizza2007reaction,vespignani2011modelling}, we then multiply equation \eqref{eq:statA} by $p(k)$ and sum over $k$ to obtain
\begin{equation}
	\bar{\rho}_I = \frac{\beta}{\mu} \bar{\Gamma},
\label{eq:rhoBgamma}
\end{equation}
which enables us to simplify Eq.~\eqref{eq:stat} to
\begin{subequations}
\begin{align}
	\bar{\rho}_{S,k} &= (1-D_S)\left[\bar{\rho}_{S,k} + \mu \bar{\rho}_{I,k} - \beta \bar{\Gamma}_k\right] + \frac{D_S k}{\expec{k}} \bar{\rho}_S , \; \label{eq:statA2} \\
	\bar{\rho}_{I,k} &= (1-D_I)\left[(1-\mu)\bar{\rho}_{I,k} + \beta \bar{\Gamma}_k\right] + \frac{D_I k}{\expec{k}} \bar{\rho}_I \; . \label{eq:statB2}
\end{align}
\label{eq:stat2}
\end{subequations}
An obvious solution to Eq.~\eqref{eq:stat2} is the trivial solution $\bar{\rho}_{S,k}=\rho$ and $\bar{\rho}_{I,k}=0$.  In the following we will mainly be interested  in non-trivial solutions.

Equations~\eqref{eq:stat2}  are valid for a general form of the reaction kernel $\bar \Gamma_k$ and have previously been derived in Refs.~\cite{colizza2007reaction,vespignani2011modelling} where it was assumed that the reactions on \emph{all} nodes of the network were  either of ``small village type'' ($\rho\ll\rho_\times$) or ``big city type'' ($\rho\gg\rho_\times$)  (and referred to as reaction type $1$ and $2$, respectively). 
However, the results from Refs.~\cite{colizza2007reaction,vespignani2011modelling} do not apply to a \emph{mixed} reaction type of the form \eqref{eq:Gamma_k} that, depending on the nodal  population density, interpolates between  type $1$ and $2$ reactions. Such dependence on the local density introduces heterogeneous dynamics in the network. In the following, we investigate the effect on the epidemic threshold from this more general reaction type, which we find more adequate for realistic systems.

%However, the results from Refs.~\cite{colizza2007reaction,vespignani2011modelling} do not apply to a \emph{mixed} reaction type of the form \eqref{eq:Gamma_k} studied here, which depending on the nodal  population density interpolates between these  limiting cases. Such dependence on local density introduces heterogeneous dynamics in the network. In the following, we investigate the effect of this more general reaction type, which we find more adequate for realistic systems, on the epidemic threshold.

\subsection{Two special cases: migrating and non-migrating susceptible individuals}

We will now consider two special cases, namely the situations where the susceptible individuals \emph{can} or \emph{cannot} diffuse while the infected individuals always are assumed to diffuse ($D_I=1$). For these two cases we will for simplicity consider  $D_S=0$ and $D_S=1$.

%
% D_S=0
%
\medskip
\emph{Case 1: Non-migrating susceptible individuals ($D_S=0$)}. 
When the susceptible individuals \emph{cannot} migrate within the metapopulation, it follows  from  Eqs.~\eqref{eq:statA2} that  \begin{align}
  \label{eq:bar_rho_I_DA0_DB1}
  \bar{\rho}_{I,k} = \frac{\beta}{\mu} \ol{\Gamma}_k = \frac{\beta}{\mu} \frac{\ol{\rho}_{S,k} \ol{\rho}_{I,k}}{\rho_\times + \ol{\rho}_{S,k} + \ol{\rho}_{I,k}}.
\end{align}
In the last transition of Eq.~\eqref{eq:bar_rho_I_DA0_DB1} we have taken advantage of the definition~\eqref{eq:Gamma_k} for the reaction kernel. By  multiplying Eq.~\eqref{eq:bar_rho_I_DA0_DB1} by $(\rho_\times + \ol{\rho}_{S,k} + \ol{\rho}_{I,k})/\ol{\rho}_{I,k}$ and  $p(k)$ (the degree distribution), summing over $k$, and using that $\rho=\bar{\rho}_S+\bar{\rho}_I$ (Eq.~\eqref{eq:rho}), leads to 
%$\ol{\rho}_S = (\mu/\beta) (\rho_\times + \rho)$
%\begin{subequations}
% \label{eq:stationary_state_case_1}
 % \begin{align}
 %   \label{eq:stationary_state_case_1_EqA}
 %   \ol{\rho}_S &= \frac{\mu}{\beta} (\rho_\times + \rho).
%  \end{align}
%In obtaining Eq.~\eqref{eq:stationary_state_case_1_EqA} it has been used that $\rho=\bar{\rho}_S+\bar{\rho}_I$ [see Eq.~\eqref{eq:rho}], from which one may also deduce that 
\begin{align}
  \ol{\rho}_I = \rho \left(1- \frac{\mu}{\beta}\right) - \frac{\mu}{\beta}\rho_\times,  \qquad D_S=0.
  \label{eq:stat4M}
\end{align}
Formally this expression is only valid when both $\ol{\rho}_S\geq 0$ and $\ol{\rho}_I\geq 0$. Moreover, since the parameters entering this equation are non-negative, it follows that one must also require $\beta/\mu>1$. A critical nodal density, $\rho=\rho_c$, is found by solving Eq.~\eqref{eq:stat4M} under the assumption of  $\ol{\rho}_I = 0$. This leads us to write the stationary state of the nodal density of infected individuals in the form 
\begin{subequations}
  \label{eq:CriticalPoint_DA_0}
\begin{align}
  \ol{\rho}_I =& 
       \begin{cases}
         \rho - \rho_c, & \rho > \rho_c\\
         0,             & \rho \leq \rho_c
       \end{cases},   
       \label{eq:CriticalPoint_DA_0_EqA}
\end{align}
where the critical nodal density reads
\begin{align}
  \label{eq:critical_density_Da=0}
  %\rho_c = \frac{\mu}{\beta-\mu}\rho_\times, \qquad D_S=0.
  \rho_c = \displaystyle\frac{\rho_\times}{\frac{\beta}{\mu}-1}, \qquad D_S=0,
\end{align}
\end{subequations}
which is the same result as for an isolated population (see Eq.~\eqref{eq:epidemic_treshold}). This means that when the susceptible individuals are immobile the epidemic threshold is independent of the network's topology.
%Since $\rho_c$, $\rho_\times$, $\beta$ and $\mu$ all are non-negative quantities, it follows that one must require $\beta/\mu>1$. Notice from Eq.~\eqref{eq:critical_density_Da=0} that $\rho_c/\rho_\times$ only depends on the infection and recovery rates, $\beta$ and $\mu$, respectively,  and is in particular independent of the nodal density $\rho$. Below we will see that when $D_S=1$ the same ratio will in fact show an explicit dependence on $\rho$. 

%
% D_S=1 
%
\medskip
\emph{Case 2: Migrating susceptible individuals ($D_S=1$)}. 
We will now address the situation where both susceptible and infected individuals are allowed to migrate between subpopulations. Under the assumption that $D_S=D_I=1$, it follows from Eq.~\eqref{eq:stat2} that 
\begin{align}
  \label{eq:stat_density_k}
  \ol{\rho}_{S,k} = \frac{k}{\left<k\right>}\ol{\rho}_S; \qquad   \ol{\rho}_{I,k} = \frac{k}{\left<k\right>}\ol{\rho}_I,
\end{align}
where the nodal densities of both infected and uninfected individuals are proportional to the node degree. The same result is also found for  pure diffusion processes where no infection takes place at all~\cite{SimonsenDiff,SimonsenDiff1}. By introducing relations~\eqref{eq:stat_density_k} into Eq.~\eqref{eq:rhoBgamma}, solving the resulting equation for $\ol{\rho}_{S}$ and using $\rho= \ol{\rho}_{S}+\ol{\rho}_{I}$, we obtain
\begin{align}
  \label{eq:CriticalPoint_DA_1}
  \ol{\rho}_I 
     &= \rho - \frac{\rho_\times}{\beta/\mu} 
            \left[ \sum_k  \frac{ 
                                  p(k) \frac{ k^2 }{ \left<k\right>^2 } 
                               }{
                                  1 +   \frac{ k }{ \left<k\right> } \frac{ \rho }{\rho_\times }
                                }
            \right]^{-1},
       \qquad D_S=1.
\end{align}
This is valid when $0 < \ol{\rho}_I \leq \rho$ and  $\ol{\rho}_I =0$ otherwise. Contrary to the $D_S=0$ case, Eq.~\eqref{eq:CriticalPoint_DA_1} shows that $\ol{\rho}_I$ is a complicated function depending both on \emph{local} interaction parameters ($\beta$, $\mu$ and $\rho_\times$) and \emph{global} network parameters~($p(k)$) when susceptible individuals are allowed to migrate. 
By taking the ``small village'' and ``big city'' limits of Eq.~\eqref{eq:CriticalPoint_DA_1} one obtains 
\begin{align}
  \label{eq:CriticalDensity_DA_1_Expansion}
  \ol{\rho}_I 
     =& 
       \begin{cases}
         \rho - \frac{\rho_\times}{\beta/\mu} \frac{\left< k \right>^2}{\left< k^2 \right>}, 
             &  \frac{k_m }{\left< k \right>} \frac{\rho}{\rho_\times} \ll 1 
         \\
         \rho \left( 1- \frac{1}{\beta/\mu} \right),                      
             &  \frac{k_m }{\left< k \right>} \frac{\rho}{\rho_\times} \gg 1 
       \end{cases},
\end{align}
where $k_m$ denotes the maximum degree of the underlying network (note that $\beta/\mu>1$). Notice that the quantity $\frac{k_m }{\left< k \right>} \frac{\rho}{\rho_\times}$ is not only large in the ``big city'' limit where $\rho/\rho_\times\gg 1$, but potentially also in the ``small village'' limit ($\rho/\rho_\times\ll 1$) if the network is sufficiently heterogeneous so $k_m/\left< k \right>$ is large enough.

The epidemic threshold, $\rho=\rho_c$, is obtained by setting $\ol{\rho}_I=0$ in Eq.~\eqref{eq:CriticalPoint_DA_1}.  In this case, however, we cannot get $\rho_c$ in closed form. Rather, it is defined implicitly through 
\begin{align}
  \label{eq:CriticalDensity_DA_1}
    \frac{\beta}{\mu} \frac{\rho_c}{\rho_\times}
            \left[ \sum_k  \frac{ 
                                  p(k) \frac{ k^2 }{ \left<k\right>^2 }
                               }{
                                  1 +   \frac{ k }{ \left<k\right> } \frac{ \rho_c }{\rho_\times }
                                }
            \right] =& 1.
       %\qquad D_S=1.
\end{align}
Since the summation cannot be evaluated explicitly for a general degree-distribution $p(k)$, one must resort to numerical methods to calculate $\rho_c$ for a given set of parameters. Exceptions do however exist. One simple example is a regular grid, \ie a homogeneous network,  where  \emph{every} node has $\left<k \right>=k$ links. In this case Eq.~\eqref{eq:CriticalDensity_DA_1} yields the same critical density as when the susceptible individuals are immobile (Eq.~\eqref{eq:critical_density_Da=0}), for which there is no dependence on the size of the network ($V$) or on the number of links each node has (as long as all nodes have the same number of links).

Equations~\eqref{eq:CriticalPoint_DA_0} and \eqref{eq:CriticalPoint_DA_1} are analytic mean-field expressions for the  nodal density of infected individuals in the stationary state. In Fig.~\ref{fig:typeM} these expressions are compared to simulations (discussed in Sec. \ref{sec:numerical_results}) for a few model parameters; They corroborate well with each other%
\footnote{The agreement between the simulations and mean-field results found in this work is better than what has been reported previously in Ref.~\cite{colizza2007reaction,vespignani2011modelling}. This in particular applies to the position of the epidemic threshold and the value of $\ol{\rho}_I$ for $\rho>\rho_c$. We speculate that the poorer agreement found in Ref.~\cite{colizza2007reaction,vespignani2011modelling} is caused by too high values for the rates $\beta$ and $\mu$ being used in the simulations. Since these rates are not explicitly given in Ref.~\cite{colizza2007reaction,vespignani2011modelling} we base our speculation on being able to reproduce their reported behavior by increasing the values of the rates.}. 
In particular we point out that the epidemic threshold, the point at which $\ol{\rho}_I$ goes from zero to being non-zero, is well described. Even if the agreement between the simulation and analytic results is satisfactory, it should be stressed that the networks used in the simulations were only characterized by their degree distributions and were otherwise random. Specifically, no clustering or community structures were present~\cite{SantoCommunities} and it is left for future research to investigate how well the mean-field results apply to such cases.

\subsection{Epidemic threshold dependence on model parameters}

% --- Figure (start) -----------------------------------------------------

\begin{figure}[tbp] %[ht]=here [t]=top [b]=bottom [p]=separat side
\centering
\includegraphics[width=0.65\textwidth]{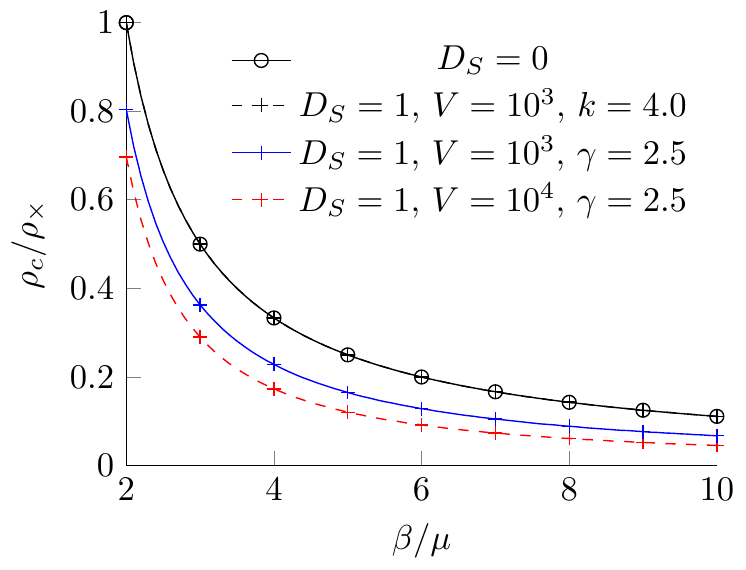}  % Putter inn eps-fila
%
%\subfigure[$\rho_c/\rho_\times$ {\it vs} $\beta/\mu$ ]{
%\includegraphics[width=0.45\textwidth]{statTM}  % Putter inn eps-fila
%\label{fig:rho_critical-A}
%}
%
%\subfigure[$\rho_c/\rho_\times$ {\it vs} $\rho/\rho_\times$]{
%\includegraphics[width=0.45\textwidth]{statTM}  % Putter inn eps-fila
%\label{fig:rho_critical-B}
%}
%
\caption[]{The epidemic threshold $\rho_c$ (scaled by $\rho_\times$), {\it vs.} the ratio of the infection and recovery rates, $\beta/\mu$,  for various networks and sizes~($V$). Equations~\eqref{eq:critical_density_Da=0} or \eqref{eq:CriticalDensity_DA_1} were used to  calculate $\rho_c$ when $D_S=0$ or $D_S=1$, respectively. The curve labeled by $k$ corresponds to a regular grid where all nodes have degree $k$. Likewise, the lines labeled by $\gamma$ correspond to a scale-free degree distribution $p(k)\sim k^{-\gamma}$. For all curves it was assumed that the infected individuals were mobile with $D_I=1$. 
}
\label{fig:rho_critical}
\end{figure}

% --- Figure (end) -----------------------------------------------------

In the previous section we established that the epidemic threshold $\rho_c$ is well described by the mean-field results Eqs.~\eqref{eq:critical_density_Da=0} and \eqref{eq:CriticalDensity_DA_1}. Here we will discuss in more detail how $\rho_c$ depends on the model parameters. In Fig.~\ref{fig:rho_critical} we present $\rho_c/\rho_\times$ as a function of the rate ratio $\beta/\mu$ for migrating ($D_S=1$) and non-migrating ($D_S=0$) susceptibles for a few different networks. The general trend is that the threshold drops as the infection rate $\beta$ becomes much larger than the recovery rate $\mu$. Moreover, when $D_S=1$ (or more precisely $D_S>0$), the epidemic threshold also drops when the degree distribution of the network becomes heterogeneous~(fat-tailed), or when the size of the network increases.  

The two overlapping black curves in Fig.~\ref{fig:rho_critical} correspond to either immobile susceptibles~($D_S=0$) on any network type, or mobile susceptibles~($D_S=1$) on a regular grid where all nodes have $k=4$ neighbours.  The epidemic thresholds for these cases are given by the simple expression $\rho_c/\rho_\times = (\beta/\mu-1)^{-1}$ (see Eq.~\eqref{eq:critical_density_Da=0}), and have no dependence on network size.

The two lower curves in Fig.~\ref{fig:rho_critical} show $\rho_c$ for two scale-free networks ($p(k)\sim k^{-2.5}$) of different size. Within a scale-free metapopulation the local population density varies quite dramatically which means that the dynamics in some nodes are of the ``small village'' type and some (typically much fewer) are of the ``big city'' type, and the rest somewhere in between. As the network gets bigger, or the ``small village'' dynamics dominate ($\rho_\times\gg \ol\rho_k$ for most nodes), the epidemic threshold drops. This behavior can be deduced analytically from Eq.~\eqref{eq:CriticalDensity_DA_1} (or Eq.~\eqref{eq:CriticalDensity_DA_1_Expansion}) to give 
\footnote{In practice we let $\frac{k_m }{\left< k \right>} \frac{\rho}{\rho_\times} \ll 1$, where $k_m$ is the maximum number of links, in Eq.~\eqref{eq:CriticalDensity_DA_1}. }
\begin{align}\label{eq:CriticalDensity_limit}
  \frac{\rho_c}{\rho_\times} \sim& \frac{1}{\beta/\mu} \frac{\left< k \right>^2}{\left< k^2 \right>}.
  % \qquad \frac{k_m }{\left< k \right>} \frac{\rho}{\rho_\times} \ll 1.
\end{align}
Here we see directly that for a very heterogeneous degree distribution or an infinite network ($V\rightarrow \infty$), both of which yields $\left<k^2 \right> \gg \left<k \right>^2$, the epidemic threshold tends to zero when keeping $\beta/\mu$ finite.

From the general result Eq.~\eqref{eq:CriticalDensity_DA_1} we can also deduce the relationship between $\rho_c$ and $\rho_\times$. Since this equation is a function of the fraction  $\rho_c/\rho_\times$, we conclude that they are proportional to each other.  The proportionality constant is $ (\beta/\mu-1)^{-1}$   (see Eq.~\eqref{eq:critical_density_Da=0}) for immobile susceptibles ($D_S=0$) and for a regular grid ({\it e.g.} $k=4$ neighbours). For large and heterogenous networks, or when ``small city'' dynamics dominate, it is given by  Eq. \eqref{eq:CriticalDensity_limit}.

% % --- Figure (start) -----------------------------------------------------

% \begin{figure}[tbp] %[ht]=here [t]=top [b]=bottom [p]=separat side
% \centering
% %
% %
% \subfigure[$\rho_c/\rho_\times$ {\it vs} $\rho/\rho_\times$ for various networks types]{
% \includegraphics[width=0.45\textwidth]{statTM}  % Putter inn eps-fila
% \label{fig:rho_critical_network_types-A}
% }
% %
% \caption[]{To be added}
% %
% \label{fig:rho_critical_network_types}
% \end{figure}

% % --- Figure (end) -----------------------------------------------------

\section{Summary and conclusions}
\label{Sec:Conclusions}

In this work we have reviewed and applied a generalized reaction-diffusion approach to epidemic spreading within a metapopulation with the overall goal to study the impact of city-size heterogeneity. As  the epidemic model we assumed the well-studied susceptible-infected-susceptible~(SIS) system. This model exhibits an epidemic and a non-epidemic phase, of which the latter has a nonzero number of infected individuals in the long time limit.  The transition between these states, the so-called epidemic threshold, depends non-trivially on the parameters of the model. These parameters are the infection and recovery rates, the cross-over population size separating  small village dynamics from large city dynamics, and the topology of the network distributing the subpopulations within the metapopulation. We envision the network to symbolize how cities and villages are interconnected in a transportation network, and human traveling patterns are assumed as simple as possible; People diffuse randomly. In principle, the diffusion rates also  enter into the epidemic threshold, but here we only addressed either immobile and mobile susceptibles ($D_S=0$ or $D_S=1$). Instead,  we paid special attention to a previously unexplored case, namely, city-size dependence on epidemic spreading. By city size we mean population density. It is an interesting aspect because a disease spreads quite differently in a small city (a village) compared to in a metropolitan city, and the most realistic situation is that all city sizes coexist within a metapopulation. We therefore formally let the local dynamics in each node adapt itself to the instantaneous local population density. In a diffusion model, the size of the city is proportional to the number of its links (we assume an undirected, non-weighted network). This means that the network's structural features enter the dynamics of the disease in two ways. First, how people migrate, and second, the dynamics in each node. Previous works have assumed that the individuals are infected in the same way in all nodes. Our main result is that we obtain (in the mean-field limit) and 
for a general network an implicit expression for the epidemic threshold expressed in terms of the  model parameters. This expression shows good agreement with numerical simulations and its asymptotic behaviour interpolates between known limits.

Epidemic spreading is a topic of great social importance that also has attracted the interest of diverse scientific disciplines for decades.   The results presented in this paper open for a better understanding of the position of the epidemic threshold and its complex interplay between the local infection parameters and the global parameters characterizing the network topology for a large class of diseases (SIS).  Such knowledge may in the future prove critical for reducing the outbreak size, slowing down the speed of spreading and/or spatially confining infectious diseases. Moreover, if epidemic dynamics were better understood, our contemporary societies could earlier, and at better levels of accuracy, predict if a local outbreak has a risk of becoming a global pandemic. If an accurate early warning could be given, appropriate measures could be taken, for example by adjusting the production of vaccines,  with the potential of saving numerous lives.

\section*{Acknowledgments}

L.L acknowledges financial support from the Knut and Alice Wallenberg foundation, and 
I.S. thanks O.M. Brende for stimulation discussions on the topic of this paper.

%I.S. would like to ......
%This research of I.S. was supported in part by the Research
%Council of Norway (Sm{\aa}forsk grant).
%\end{acknowledgements}

% --------------------------------------------------------------------
% REFERENCES
% --------------------------------------------------------------------

%\bibliographystyle{unsrt}       % APS-like style for physics
\bibliographystyle{spphys} 
\bibliography{paper2010-07}

\begin{thebibliography}{10}
\providecommand{\url}[1]{{#1}}
\providecommand{\urlprefix}{URL }
\expandafter\ifx\csname urlstyle\endcsname\relax
  \providecommand{\doi}[1]{DOI \discretionary{}{}{}#1}\else
  \providecommand{\doi}{DOI \discretionary{}{}{}\begingroup
  \urlstyle{rm}\Url}\fi

\bibitem{dawood2012estimated}
F.~Dawood, A.~Iuliano, C.~Reed, M.~Meltzer, D.~Shay, P.~Cheng,
  D.~Bandaranayake, R.~Breiman, W.~Brooks, P.~Buchy, et~al., The Lancet
  Infectious Diseases  (2012)

\bibitem{who2003}
\url{http://www.who.int/csr/sars/country/2003_08_15/en/index.html}

\bibitem{ross1916application}
R.~Ross, Proc. R. Soc. London A \textbf{92}, 204 (1916)

\bibitem{hamer1929epidemiology}
W.~Hamer, \emph{Epidemiology, old and new} (The Macmillan company, 1929)

\bibitem{kermack1927contributions}
W.O. Kermack, A.G. McKendrick, Proc. R. Soc. London A \textbf{115}, 700 (1927)

\bibitem{mollison1977spatial}
D.~Mollison, J. Royal Stat. Soc. B \textbf{39}, 283 (1977)

\bibitem{rhodes1997critical}
C.~Rhodes, H.~Jensen, R.~Anderson, Proc. R. Soc. London B \textbf{264}, 1639
  (1997)

\bibitem{sneppen2010minimal}
K.~Sneppen, A.~Trusina, M.~Jensen, S.~Bornholdt, PloS one \textbf{5}, e13326
  (2010)

\bibitem{ferguson2005strategies}
N.~Ferguson, D.~Cummings, S.~Cauchemez, C.~Fraser, S.~Riley, A.~Meeyai,
  S.~Iamsirithaworn, D.~Burke, Nature \textbf{437}, 209 (2005)

\bibitem{colizza2007reaction}
V.~Colizza, R.~Pastor-Satorras, A.~Vespignani, Nature Phys. \textbf{3}, 276
  (2007)

\bibitem{vespignani2011modelling}
A.~Vespignani, Nature Phys. \textbf{8}, 32 (2011)

\bibitem{colizza2007invasion}
V.~Colizza, A.~Vespignani, Phys. Rev. Lett. \textbf{99}, 148701 (2007)

\bibitem{grenfell2001travelling}
B.~Grenfell, O.~Bjornstad, J.~Kappey, Nature \textbf{414}, 716 (2001)

\bibitem{earn2000simple}
D.~Earn, P.~Rohani, B.~Bolker, B.~Grenfell, Science \textbf{287}, 667 (2000)

\bibitem{brockmann2009human}
D.~Brockmann, Rev. Nonlin. Dyn. and Compl. \textbf{2}, 1 (2009)

\bibitem{hufnagel2004forecast}
L.~Hufnagel, D.~Brockmann, T.~Geisel, Proc. Nat. Acad. Sci. USA \textbf{101},
  15124 (2004)

\bibitem{colizza2006role}
V.~Colizza, A.~Barrat, M.~Barth{\'e}lemy, A.~Vespignani, Proc. Nat. Acad. Sci.
  USA \textbf{103}, 2015 (2006)

\bibitem{colizza2007modeling}
V.~Colizza, A.~Barrat, M.~Barthelemy, A.~Valleron, A.~Vespignani, PLoS Medicine
  \textbf{4}, e13 (2007)

\bibitem{ferguson2006strategies}
N.~Ferguson, D.~Cummings, C.~Fraser, J.~Cajka, P.~Cooley, D.~Burke, Nature
  \textbf{442}, 448 (2006)

\bibitem{brockmann2006scaling}
D.~Brockmann, L.~Hufnagel, T.~Geisel, Nature \textbf{439}, 462 (2006)

\bibitem{balcan2011phase}
D.~Balcan, A.~Vespignani, Nature Phys. \textbf{7}, 581 (2011)

\bibitem{newman2002spread}
M.E.J. Newman, Phys. Rev. E \textbf{66}, 016128 (2002)

\bibitem{pastor2001epidemic}
R.~Pastor-Satorras, A.~Vespignani, Phys. Rev. Lett. \textbf{86}, 3200 (2001)

\bibitem{kuperman2001small}
M.~Kuperman, G.~Abramson, et~al., Phys. Rev. Lett. \textbf{86}, 2909 (2001)

\bibitem{keeling2000metapopulation}
M.~Keeling, C.~Gilligan, et~al., Nature \textbf{407}, 903 (2000)

\bibitem{pinto2012locating}
P.~Pinto, P.~Thiran, M.~Vetterli, Phys. Rev. Lett. \textbf{109}, 68702 (2012)

\bibitem{Book:Keeling-2008}
M.~Keeling, P.~Rohani, \emph{Modeling Infectious Diceases in Humans and
  Animals} (Princeton University Press, 2008)

\bibitem{Newman}
M.E.J. Newman, SIAM Review \textbf{45}, 167 (2003)

\bibitem{SantoCommunities}
S.~Fortunato, Phys. Rep. \textbf{486}, 75 (2010)

\bibitem{Book:Redner-2001}
S.~Redner, \emph{A Guide to First-Passage Processes} (Cambridge University
  Press, Cambridge, 2001)

\bibitem{SimonsenInternet}
K.A. Eriksen, I.~Simonsen, S.~Maslov, K.~Sneppen, Phys. Rev. Lett. \textbf{90},
  148701 (2003)

\bibitem{SimonsenDiff}
I.~Simonsen, Physica A \textbf{357}, 317 (2005)

\bibitem{SimonsenDiff1}
I.~Simonsen, K.A. Eriksen, S.~Maslov, K.~Sneppen, Physica A \textbf{336}, 163
  (2004)

\bibitem{SimonsenCascading}
I.~Simonsen, L.~Buzna, K.~Peters, S.~Bornholdt, D.~Helbing, Phys. Rev. Lett.
  \textbf{100}, 218701 (2008)

\bibitem{Molloy}
M.~Molloy, B.~Reed, Random Structures and Algorithms \textbf{6}, 161 (1995)

\bibitem{stanley1987introduction}
H.~Stanley, \emph{Introduction to phase transitions and critical phenomena}
  (Oxford University Press, 1987)

\bibitem{Saldana}
J.~Saldana, Phys. Rev. E \textbf{78}, 012902 (2008)

\bibitem{NewmanStrogatz}
M.E.J. Newman, S.H. Strogatz, D.J. Watts, Phys. Rev. E \textbf{64}, 026118
  (2001)

\end{thebibliography}

\end{document}